\documentclass[twocolumn,floatfix,preprintnumbers,citeautoscript,newabstract,nofootinbib]{revtex4} 
\usepackage{amsmath,amssymb}
\usepackage{bbm} 
\usepackage{mathrsfs}	

\newcommand{\I}{\mathrm{i}}

\newcommand{\U}[1]{\ensuremath{\mathrm{U}(#1)}}
\newcommand{\Z}[1]{\ensuremath{\mathbbm{Z}_{#1}}} 
\DeclareMathOperator{\re}{Re}

\begin{document} \setlength{\unitlength}{1in}

\preprint{TUM-HEP-705/08; DESY 08-189; LMU-ASC 60/08\\[0.5cm] }

\title{\Large Large hierarchies from approximate $\boldsymbol{R}$ symmetries}

\author{{\bf
Rolf~Kappl$^{1}$\!,
Hans Peter~Nilles$^2$\!, 
Sa\'ul~Ramos-S\'anchez$^3$\!, 
Michael~Ratz$^1$\!,}\\{\bf
Kai~Schmidt-Hoberg$^1$\!, Patrick~K.~S.~Vaudrevange$^4$
}\\[0.5cm]
{\it
${}^1$ Physik Department T30, Technische Universit\"at M\"unchen,}\\[-0.05cm]
{\it\normalsize James-Franck-Strasse, 85748 Garching,
Germany}\\[0.15cm]
{\it
${}^2$ Bethe Center for Theoretical Physics and Physikalisches Institut der Universit\"at Bonn,}\\[-0.05cm]
{\it
Nussallee 12, 53115 Bonn,
Germany}\\[0.15cm]
{\it
${}^3$ Deutsches Elektronen-Synchrotron DESY, Notkestrasse 85, 22603 Hamburg,
Germany}\\[0.15cm]
{\it
${}^4$ Arnold Sommerfeld Center for Theoretical Physics,}\\[-0.05cm]
{\it
Ludwig-Maximilians-Universit\"at M\"unchen, 80333 M\"unchen,
Germany}
}

\begin{abstract}
We show that hierarchically small vacuum expectation values of the
superpotential in supersymmetric theories can be a consequence of an approximate
$R$ symmetry. We briefly discuss the role of such small constants in moduli
stabilization and understanding the huge hierarchy between the Planck and
electroweak scales.
\end{abstract}

\pacs{\dots}

\maketitle

\section{Introduction}

One of the major puzzles in contemporary physics is the existence of large
hierarchies in nature, such as the ratio between the Planck and electroweak
scales $M_\mathrm{P}/m_W\sim 10^{17}$. Some of the most promising explanations
of such hierarchies rely on dimensional transmutation. Here the dynamical scale
$\Lambda=M_\mathrm{P}\,\mathrm{e}^{-a/g^2}$ (with $g$ and $a$ denoting the gauge
coupling and a constant, respectively) can be naturally much smaller than the
fundamental scale. However, if one is to embed this mechanism in a more
fundamental framework, one often encounters the problem that there has to be a
hierarchically small quantity right from the start. Concretely, if one is to
make use of the dynamical scale in string theory, one has first to fix the
modulus that determines the coupling strength. This in turn often requires the
introduction of a small constant. One faces then the well-known
``chicken-or-egg problem''.

Motivated by results obtained in the framework of string theory model
building, we present here a potential resolution of the
problem. We shall show that, if the superpotential in a supersymmetric theory
exhibits an approximate $\U1_R$ symmetry, it generically acquires a suppressed
vacuum expectation value (VEV).  Such accidental $\U1_R$ symmetries which get
broken at higher orders are naturally present in string compactifications.  They
arise as remnants from exact, discrete $R$ symmetries. Such symmetries allow us
to control the VEV of the (perturbative) superpotential and, in particular, to
avoid deep anti-de Sitter vacua.   We will discuss the role of the resulting
hierarchically small superpotential VEVs in the context of moduli stabilization
in string theory, for giving a plausible explanation of the huge hierarchy
between $M_\mathrm{P}$ and $m_W$, and for providing, in the context of a class
of promising string models~\cite{Lebedev:2007hv}, a solution to the $\mu$ problem of the minimal
supersymmetric standard model (MSSM).

\section{Supersymmetric Minkowski Vacua as a consequence of a $\boldsymbol{\U1_R}$
symmetry}
\label{sec:ExactR}

Consider a superpotential of the form
\begin{equation}\label{eq:Wgeneral}
 \mathscr{W}~=~\sum c_{n_1\cdots n_M} \phi_1^{n_1}\cdots \phi_M^{n_M}\;.
\end{equation} 
Here and in the following we work in Planck units, i.e.\ we set
$M_\mathrm{P}=1$ unless stated differently.
Assume that $\mathscr{W}$ has an exact $R$ symmetry, under which $\mathscr{W}$
has $R$ charge 2,
\begin{equation}\label{eq:U1RW}
 \mathscr{W}~\to~\mathrm{e}^{2\I\,\alpha}\,\mathscr{W}\;,
\end{equation}
and the fields transform as
\begin{equation}
 \phi_j~\to~\phi_j'~=~\mathrm{e}^{\I\,r_j\,\alpha}\,\phi_j
\end{equation}
such that each monomial in \eqref{eq:Wgeneral} has total $R$ charge 2.

Let $\langle\phi_i\rangle$ denote a field configuration which solves the $F$-term
equations,
\begin{equation}
 F_i~=~\frac{\partial\mathscr{W}}{\partial\phi_i}~=~0
 \quad \text{at}~\phi_j=\langle\phi_j\rangle~\forall~i,j\;.
\end{equation}
Consider now an infinitesimal $\U1_R$ transformation,
\begin{equation}
 \mathscr{W}(\phi_i)~\to~
 \mathscr{W}(\phi_i')
 ~=~\mathscr{W}(\phi_i)+\sum_j \frac{\partial\mathscr{W}}{\partial\phi_j}
 \Delta\phi_j\;.
\end{equation}
At $\phi_j=\langle\phi_j\rangle$ the superpotential goes into itself, which can only
be consistent with \eqref{eq:U1RW} if $\mathscr{W}=0$ at
$\phi_j=\langle\phi_j\rangle$. This proves that, if the $F$ equations are satisfied,
$\mathscr{W}$ vanishes.

A few comments are in order. First, this statement holds regardless of whether
the configuration $\langle\phi_i\rangle$ preserves $\U1_R$ or breaks it
spontaneously. Second, in the context of supergravity, the statements above
imply that the $D_i\mathscr{W}$ vanish for $\phi_i=\langle\phi_i\rangle$, i.e.\ also
the supergravity $F$ terms vanish and one obtains a supersymmetric Minkowski
vacuum. Third, our findings are related to an observation by Nelson and Seiberg 
made in \cite{Nelson:1993nf}, where it is stated that, in order to
have a theory without supersymmetric ground state, the superpotential has to
exhibit a continuous $R$ symmetry. The statements do, however, not tell us
whether or not a theory with a superpotential exhibiting a continuous $R$
symmetry has a supersymmetric ground state or not.  Our findings and
\cite{Nelson:1993nf} imply that, if there is a continuous $R$ symmetry, there are
two options:
\begin{enumerate}
 \item there is a supersymmetric ground state with $\mathscr{W}=0$
 (with $\U1_R$ spontaneously broken or unbroken);
 \label{caseSUSY}
 \item there is no supersymmetric ground state, and in the ground state $\U1_R$
 is spontaneously broken~\cite{Nelson:1993nf}.\label{caseNoSUSY}
\end{enumerate}

In this letter we focus on case~\ref{caseSUSY}. 
If the \U1 that acts on the scalar components of the superfields gets
spontaneously broken at $\phi_i=\langle\phi_i\rangle$ (which is the case if, for
instance, all $\langle\phi_i\rangle$ are non-trivial),  it follows then from
Goldstone's theorem that there is a massless mode, the so-called $R$ axion.

\section{Small constants from approximate $\boldsymbol{\U1_R}$ symmetries}
\label{sec:SmallConstants}

Let us now study what happens if the $R$ symmetry is `slightly' broken, i.e.\ by
higher order terms. We can write the superpotential as
\begin{equation}
 \mathscr{W}(\phi_i)~=~  \mathscr{W}_0(\phi_i) + \sum_j  \mathscr{W}_j(\phi_i)
 \;,
\end{equation}
where $\mathscr{W}_0(\phi_i)$ consists of monomials up to order $N-1$ which 
preserve the $R$ symmetry while the $\mathscr{W}_j(\phi_i)$
are monomials of order $\ge N$ which break the $R$ symmetry.
This means that the superpotential transforms under $\U1_R$ as
\begin{align}
 \mathscr{W}(\phi_i) &\rightarrow~ \mathrm{e}^{2\I \alpha}\mathscr{W}_0(\phi_i) 
                     + \sum_j \mathrm{e}^{ \I \alpha\,R_j} \mathscr{W}_j(\phi_i) 
\nonumber \\
& \simeq~  \mathscr{W}(\phi_i) + \I\, \alpha\, \left(2\mathscr{W}_0(\phi_i) 
          + \sum_j R_j \,\mathscr{W}_j(\phi_i) \right) 
\end{align}
with $R_j\ne 2$, and
\begin{align}
 \mathscr{W}(\phi_i) & \rightarrow~ \mathscr{W}(e^{\I \alpha\,r_i} \phi_i)
\nonumber \\
& \simeq~  \mathscr{W}(\phi_i) 
+ \I\, \alpha\, \sum_j \frac{\partial \mathscr{W}}{\partial \phi_j}
r_j\, \phi_j \;.
\end{align}
Combining these two expressions and assuming that the $F$-terms vanish in our vacuum,
$\frac{\partial \mathscr{W}}{\partial \phi_i}=0$, we see that
\begin{equation}\label{eq:NonTrivialW}
 \langle \mathscr{W}(\phi_i) \rangle
 ~=~
 -\frac{1}{2}\sum_j (R_j-2)\langle\mathscr{W}_j(\phi_i) \rangle \;.
\end{equation}
This means that in the case of an approximate $\U1_R$ symmetry one obtains
suppressed superpotential VEVs, written symbolically as
\begin{equation}
 \langle\mathscr{W}\rangle~\sim~\langle \phi\rangle^{\ge N}\;.
\end{equation}

In many situations there is a mild hierarchy between the fundamental scale and a
typical VEV, $\langle \phi\rangle/M_\mathrm{P}<1$. This is,
for instance, the case in string models where a \U1 factor appears `anomalous',
and where the one-loop Fayet-Iliopoulos term forces some VEVs to be roughly one order
of magnitude smaller than $M_\mathrm{P}$~\cite{Dine:1987xk}. According to the
above discussion, the suppression of $\langle\mathscr{W}\rangle$ gets then
enhanced by the $N^\mathrm{th}$ power of this mild hierarchy, similarly to
the Froggatt-Nielsen picture~\cite{Froggatt:1978nt}.

Further, we have seen that there might be a Goldstone mode $\eta$. 
With explicit $\U1_R$ breaking, it will generically receive a mass,
$m_\eta\sim\langle \phi\rangle^{\ge N-2}$. 
(The ``$-2$'' comes from the second derivative.)
In supergravity theories, $\langle\mathscr{W}\rangle$ sets the gravitino mass,
\begin{equation}
 m_{3/2}~\simeq~\langle\mathscr{W}\rangle\;.
\end{equation}
This leads then to the expectation that there is a mode whose
(supersymmetric) mass scales like $m_{3/2}$,
\begin{equation}\label{eq:meta}
 m_\eta~\sim~\frac{m_{3/2}}{\langle\phi\rangle^2}\;.
\end{equation}

Let us comment that, if one is to include supergravity effects,
$\mathscr{W}\ne0$ does not necessarily imply anti-de Sitter solutions (see e.g.\
the discussion in~\cite[section~4]{Weinberg:1988cp}).

\section{Explicit string theory realization}
\label{sec:Applications}

One of the central themes of string theory is the issue of moduli stabilization,
which is closely connected to the question of supersymmetry breaking. In the
traditional approach, supersymmetry is broken by dimensional transmutation
\cite{Witten:1981nf}, e.g.\ by gaugino
condensation~\cite{Nilles:1982ik}. 
However, for this elegant mechanism to work, one needs first to fix the gauge
coupling, whose strength is given by the VEV of the dilaton $S$ or
another modulus in string theory. This can be achieved in various ways: for
instance, in the race-track scheme \cite{Krasnikov:1987jj} one has two competing
non-perturbative superpotentials which provide a non-trivial minimum of the dilaton potential.
The drawback of this mechanism is that it only works if one has two rather large
`hidden' gauge groups with rather special matter contents. A somewhat more
economic scheme is that of K\"{a}hler stabilization
\cite{Casas:1996zi,Binetruy:1996xj} where one needs only one hidden sector.
However, in the relevant regime where dilaton stabilization may be achieved the
theory is not calculable. More recently, an alternative has been studied 
 (with the most prominent example being that of KKLT \cite{Kachru:2003aw})
where the superpotential is of the form
\begin{equation}\label{eq:WKKLT}
 \mathscr{W}~=~c+A\,\mathrm{e}^{-a\,S}\;.
\end{equation}
The first term $c$ is a constant and the second term reflects hidden sector
strong dynamics, i.e.\ $S$ is related to the gauge coupling, $\re S\propto 1/g^2$,
and $a$ is related to the $\beta$-function of the hidden gauge group.
In the KKLT setup, the constant comes from fluxes. The minimum of the scalar
potential for $S$ occurs at a point where
\begin{equation}\label{eq:KKLTmin}
 |a\,S\,A\,\mathrm{e}^{-a\,S} |~\sim~|c|\;.
\end{equation}
The VEV of $\mathscr{W}$, i.e.\ the gravitino mass, is of
the same order. In order to have MSSM superpartner masses at the TeV scale, the
gravitino mass cannot exceed $\mathcal{O}(100)\,\mathrm{TeV}$, hence
\begin{equation}
 |c|~\lesssim~10^{-12}
\end{equation}
in Planck units.
The small scale in this setting is therefore \emph{not} explained by dimensional
transmutation but originates from the smallness of the constant $c$. KKLT and
others argue that, due to the large number of vacua, some of them might have such
$c$ by accident.

In what follows, we will exploit the observation of
section~\ref{sec:SmallConstants} that small VEVs of the (perturbative)
superpotential can be explained by an approximate $\U1_R$ symmetry. We will use
this in order to discuss moduli stabilization in the context of the heterotic
string. 
We focus on orbifold  compactifications~\cite{Dixon:1985jw} 
since they possess many (and well-understood) discrete symmetries, which, as 
it turns out, imply approximate $\U1_R$ symmetries of the superpotentials
describing the effective field theories derived from these constructions. 
As we shall see, superpotential VEVs of the order $10^{-\mathcal{O}(10)}$ can
naturally be obtained.
Orbifold compactifications allow us to embed the MSSM into string theory
[\citealp{Buchmuller:2005jr},\citealp{Buchmuller:2006ik},\citealp{Lebedev:2007hv}].

In our calculations we focus on the models of the `heterotic MiniLandscape'
[\citealp{Lebedev:2006kn},\citealp{Lebedev:2007hv}]. These models exhibit the
standard model gauge group and the chiral matter content of the MSSM. They are
based on the \Z6-II orbifold with three factorizable tori (see
[\citealp{Kobayashi:2004ya},\citealp{Buchmuller:2006ik}] for details). The
discrete symmetry of the geometry leads to a large number of discrete symmetries
governing the couplings of the effective field theory
\cite{Hamidi:1986vh,Dixon:1986qv} (cf.\
also~[\citealp{Kobayashi:2004ya},\citealp{Buchmuller:2006ik},\citealp{Kobayashi:2006wq}]).
Apart from various bosonic discrete symmetries, one has a
\begin{equation}\label{eq:DiscreteR}
 [\Z6\times\Z3\times\Z2]_R
\end{equation}
symmetry; other orbifolds have similar discrete symmetries.
Further, in almost all of the MiniLandscape models there is, at one-loop,
a Fayet-Iliopoulos (FI) $D$-term,
\begin{equation}
 V_D~\supset~g^2\,\left(\sum_iq_i\,|\phi_i|^2+\xi\right)^2\;,
\end{equation}
where the $q_i$ denote the charges under the so-called `anomalous \U1{}'.
It turns out that, in all models with non-vanishing FI term, $\xi$ is of order
$0.1$ (see \cite{Buchmuller:2006ik} for an explicit example).
The first step of our analysis is to identify a set of standard model singlets
$\phi_i$ with the following properties:
\begin{itemize}
 \item giving VEVs to the $\phi_i$ allows us to cancel the FI
 term;
 \item there is no other field that is singlet under the gauge symmetries left
 unbroken by the $\phi_i$ VEVs.
\end{itemize}
These properties ensure that the $\langle\phi_i\rangle$ can be consistent with a
vanishing $D$-term potential and that the $F$-terms of all other massless modes
vanish, implying that it is sufficient to derive the superpotential terms
involving only the $\phi_i$ fields.  A crucial property of these superpotentials
is that they exhibit accidental $\U1_R$ symmetries that get only broken at
rather high orders $N$. As discussed, this can be regarded as a consequence of high-power
discrete $R$ symmetries (equation~\eqref{eq:DiscreteR}). $N$ depends on the
chosen $\phi_i$ configuration; as a general rule we find that the more $\phi_i$
fields are considered, the lower $N$ values emerge. For instance, in a model
where only seven fields are considered, we obtain $N=26$, on the other hand, in
the model~1 of \cite{Lebedev:2007hv} with 24 fields switched on, $\U1_R$ gets
broken at order 9. 

Given non-trivial solutions to the $F$-term equations,
\begin{equation}
 \phi_i\,\frac{\partial\mathscr{W}}{\partial \phi_i}~=~0\;,
 \quad\text{with}~\phi_i\ne0\;,
\end{equation}
one can use complexified gauge transformations to ensure vanishing $D$-terms as
well \cite{Ovrut:1981wa}. Although $D$-term constraints do not fix the scale of
the $\langle \phi_i\rangle$ in general, the requirement to cancel the FI term
introduces the scale $\sqrt{\xi}\sim 0.3$ into the problem. We search for
solutions of $V_D=V_F=0$ in the regime $|\phi_i|<1$, and find that they exist.
We explicitly verify that for such solutions the superpotential is
hierarchically small, $\langle\mathscr{W}\rangle\sim \langle \phi\rangle^N$,
where $\langle \phi\rangle$ denotes the typical size of a VEV. A very important
property of many of these configurations is that all fields acquire
(supersymmetric) masses. Hereby typically only one field has a mass of the order
$m_\eta$ (see equation~\eqref{eq:meta})
while the others are much heavier. We have also checked that these features are
robust under supergravity corrections.

Altogether we find that in the models under consideration one obtains
isolated supersymmetric field configurations with $|\phi_i|<1$ where the
VEV of the perturbative superpotential $\langle\mathscr{W}\rangle$ is 
hierarchically small. 

Before discussing applications, let us compare our findings to other recent
results \cite{Buchmuller:2008uq}. There, using the stringy selection rules,
so-called `maximal vacua' were constructed in which the superpotential vanishes
term by term (and to all orders). In our approach, each superpotential term
composed out of $\phi_i$ fields acquires a non-trivial VEV, but to the order at
which the accidental $\U1_R$ is exact, all terms cancel non-trivially. At
higher orders, a non-trivial VEV of $\mathscr{W}$ gets induced.

Let us now briefly sketch how this can be used in order to
stabilize the dilaton, whose VEV determines the gauge coupling.
After integrating out the $\phi_i$ fields,
one is left with a superpotential of the form \eqref{eq:WKKLT},
\begin{equation}\label{eq:Weff}
 \mathscr{W}_\mathrm{eff}
 ~=~ c +A\,\mathrm{e}^{-a\,S}\;,
\end{equation}
where $c=\langle\mathscr{W}\rangle=10^{-\mathcal{O}(10)}$, 
and $A\,\mathrm{e}^{-a\,S}$ describes some non-perturbative dynamics, such as
gaugino condensation
[\citealp{Nilles:1982ik},\citealp{Ferrara:1982qs},\citealp{Derendinger:1985kk},\citealp{Dine:1985rz}].
As we have discussed before in equation~\eqref{eq:KKLTmin}, this superpotential leads to
a non-trivial minimum for the dilaton.
In the MiniLandscape models, realistic
gauge couplings are correlated with favorable sizes of the dynamical scale,
$A\,\mathrm{e}^{-a\,S}/M_\mathrm{P}^2\sim\mathrm{TeV}$~\cite{Lebedev:2006tr}.
Hence, for typical expectation values
$\langle\mathscr{W}\rangle=10^{-\mathcal{O}(10)}$ one obtains reasonable gauge
couplings.
The fixing of the $T$-moduli and other issues such as `uplifting' will be
studied elsewhere. 

Another application of our findings concerns the $\mu$ term of the
MSSM. In~\cite{Casas:1992mk} it has been proposed that in models in which the
field combination $h_u\,h_d$ (with $h_u$ and $h_d$ denoting the up-type and
down-type Higgs fields, respectively) is completely neutral w.r.t.\ all
symmetries there is an interesting relation between the Higgs mass coefficient
$\mu$ and $\langle\mathscr{W}\rangle$,
\begin{equation}
 \mu~\sim~\langle\mathscr{W}\rangle\;.
\end{equation}
The heterotic MiniLandscape~\cite{Lebedev:2006kn} contains many models in which
the Higgs pair (and only the Higgs pair) has this property. Apart from the
above property, such models exhibit `gauge-top unification', i.e.\ the top
Yukawa coupling is of the order of the gauge coupling, as well as many other
desirable properties. In a  concrete example,
the benchmark model~1A of \cite{Lebedev:2007hv}, it was found that solving the
$F$-term equations for the superpotential up to order 6 always leads to
$\langle\mathscr{W}\rangle=0$. We have now obtained a better understanding of
this fact: there is a $\U1_R$ symmetry that holds up to order 11, explaining this
property. It is amazing to see that these models, constructed in order to
reproduce the MSSM spectrum and gauge interactions, exhibit so many 
appealing properties automatically.

\section{Conclusions}
\label{sec:Conclusions}

We have shown that approximate $\U1_R$ symmetries can explain the appearance of
hierarchically small constants. We find that at configurations where the
$F$-term equations are solved, the superpotential goes like
$\langle\mathscr{W}\rangle\sim\langle\phi\rangle^N$ with $\langle\phi\rangle$
denoting a typical expectation value and $N$ being the order at which $\U1_R$ gets
broken. We have analyzed various heterotic orbifold models and found that there,
due to the presence of high-power discrete $R$ symmetries, approximate $\U1_R$s
are generic. We have explicitly solved the $F$-term equations in several models,
thus obtaining points in field space in which the $F$- and $D$-term potentials
vanish, and confirmed that, for $|\phi_i|<1$, the superpotential is
hierarchically small. We have argued that such suppressed superpotential
expectation values can be the origin for the appearance of large
hierarchies in nature: they fix the scale of the gravitino mass, which in
schemes with low-energy supersymmetry sets the weak scale, and can be used to
stabilize the string theory moduli at realistic values. \\
{}\\
\textbf{Acknowledgments.}
We would like to thank F.~Br\"ummer, T.~Kobayashi and J.~Schmidt for
useful discussions.
This research was supported by the DFG cluster of excellence Origin and
Structure of the Universe, the European Union 6th framework program 
\mbox{MRTN-CT-2006-035863} "UniverseNet", LMUExcellent and the
\mbox{SFB-Transregios} 
27 "Neutrinos and Beyond" and 33 "The Dark Universe" 
by Deutsche Forschungsgemeinschaft (DFG).


\providecommand{\bysame}{\leavevmode\hbox to3em{\hrulefill}\thinspace}

\end{document}